\documentclass{llncs}
\usepackage{graphicx}
\usepackage[english]{babel}
\usepackage{afterpage}
\usepackage[section,subsection,subsubsection]{extraplaceins}
\begin{document}

\title{Contextual Query Using Bell Tests}

\author{Jo{\~{a}}o Barros, Zeno Toffano, Youssef Meguebli and Bich-Li\^{e}n Doan}
\institute{SUPELEC, 3 rue Joliot Curie, F-91192, Gif-sur-Yvette, France \email{joaocarlos.pintobarros@supelec.fr, zeno.toffano@supelec.fr,
youssef.meguebli@supelec.fr, bich-lien.doan@supelec.fr}}

\maketitle

\begin{abstract}

Tests are essential in Information Retrieval and Data
Mining in order to evaluate the effectiveness of a query. An automatic
measure tool intended to exhibit the meaning of words in context has
been developed and linked with Quantum Theory, particularly entanglement.
``Quantum like" experiments were undertaken on semantic space based
on the Hyperspace Analogue Language (HAL) method. A quantum HAL model
was implemented using state vectors issued from the HAL matrix and
query observables, testing a wide range of window sizes. The Bell
parameter $S$, associating measures on two words in a document, was
derived showing peaks for specific window sizes. The peaks show maximum
quantum violation of the Bell inequalities and are document dependent.
This new correlation measure inspired by Quantum Theory could be promising
for measuring query relevance.

\end{abstract}


\section{Introduction}

In this work we present original ``quantum-like" tests that could be
useful in the domain of Information Retrieval (IR) and Data Mining.

Context is used to disambiguate terms. Melucci  \cite{Melucci} showed
that a query or a document can be generalized, in different contexts,
as vectors, where the likelihood of context of a set of documents
can be considered. Quantum Mechanics has been invoked to enrich
the search capabilities in IR by Rijsbergen  \cite{van Rijsbergen}
by using the mathematical formalism of the Hilbert vector space. 

Analogies between concepts issued from Quantum Theory with Information
Retrieval tools have been made by several authors. Widdows \cite{Widdows-1}
uses the quantum formalism for experiments with negation and disjunction
and Arafat \cite{Arafat} shows that user needs can be represented
by a state vector. Other analogies have been stated by Li \& Cunnigham \cite{Li.Y}
such as: ``state vector"/``objects" in a collection; ``observable"/``query",
``eigenvalues"/``relevance or not for one object"; ``probability of getting
one eigenvalue"/``relevance degree of object to a query". 

Bruza and Cole explicitly calculated eigenvectors associated to a word \cite{Bruza-Cole}
and in the field of concept representation an explicit Bell inequality violation was found \cite{Kitto-Bruza}.

\section{Bell Inequality and Bell Parameters for binary outcomes}

Entanglement, which can be made manifest through Bell inequality violations \cite{Bell}
(commonly presented in the form of the CHSH inequality \cite{CHSH})
has become a very important research trend in Physics. Several experiments
have proved the existence of entangled particles  \cite{Freedman-Clauser,Aspect}
and this fact is now widely accepted. The field has fascinated many
scientists throughout the last decades also leading to much parallel
scientific and pseudoscientific research as is well described by Keiser
in a recent book \cite{Kaiser}. Part of the attraction arises because
of the concept of ``nonlocality" of the quantum world suggesting ``spooky
action at distance" (a discussion can be found in  \cite{Khrennikov}).
Even though in general the violation of Bell inequalities demands
entanglement, higher violations of the inequalities do not necessarily
mean more entanglement. 

Quantum Information has emerged bridging physics and information science.
Though initially discovered in the context of foundations of Quantum
Mechanics, the violations of Bell inequalities referred above are
nowadays a key point in a wide range of branches of Quantum Information.
Entanglement is at the heart of this field because it is seen as a
potential ``resource" for new applications such as coding or computing
 \cite{Nielsen-Chuang}. New theorems of the kind of Bell, named no-go
theorems (for example the Kochen-Specker theorem \cite{Kochen-Specker}),
have been proposed.

In practice most experiments have used polarized photons as the famous
experiment in 1982 by Aspect et al. \cite{Aspect}. More sophisticated
set-ups are constantly being proposed and discussed  \cite{Cabello,Rossler}
very often to rule out local hidden variable models. 

Some macroscopic tests have been proposed in the form of thought experiments
or combined with yes-no questions showing also Bell inequality violations,
for example by Aerts  \cite{Aerts}.

The CHSH-Bell parameter $S_{Bell}$ for tests with two binary outcomes,
$+1$ or $-1$, can be defined as follows: 
\begin{equation}
S_{Bell}=\left|E\left(A,B\right)-E\left(A,C\right)\right|+\left|E\left(B,D\right)+E\left(C,D\right)\right|\label{eq:general-bell}
\end{equation}
where $A$, $B$, $C$ and $D$ are tests and $E\left(X,Y\right)$
stands for the expectation value of the outcome of mutual tests $X$
and $Y$.

We briefly recall some important facts about the Bell parameter.
It is easily verified that $S_{Bell}$ can never exceed $4$. More
specifically in the so called classical,
local and separable situation $S_{Bell}$ lies between $0$
and $2$. In this case, for example, we could write $E\left(X,Y\right)=E\left(X\right)E\left(Y\right)$.

The case $2\le S{}_{Bell}\le2\sqrt{2}$ can be achieved with quantum
entangled states obtained experimentally with photons. Less underlined
is the case where $S{}_{Bell}>2\sqrt{2}$ , also known as the Tsirelson’s
bound \cite{Cabello,Cirel'son }. This zone between $2\sqrt{2}$ and
$4$ is called the ``no-signaling" region. The maximum value $S_{Bell}=4$
can be attained with logical probabilistic constructions often named
PR boxes  \cite{Popescu}.

\section{Bell Tests in Semantic Space using HAL}

Our approach presented here can be perceived as an experiment done
on objects outside the domain of physics. The objects are words within
texts. We study the relationships between words within a document;
these relationships can be formed by creating a ``semantic space" using
the Hyperspace Analogue Language (HAL) method  \cite{Lund}. 

The HAL algorithm does not require any explicit human a-priori judgment.
In this procedure a HAL lexical co-occurence matrix is built with
a ``window" representing a span of words
passed over the corpus being analyzed. The width of this window can
be varied. Words within the window are recorded as co-occurring with
strength inversely proportional to the number of other words separating
them within the window. 

The point of the co-occurrence matrix is that the rows effectively
constitute vectors in a high-dimensional space, so that that the elements
of the normalized vectors are frequency counts (probabilities), and
the dimensionality of the space is determined by the number of columns
in the matrix (context vectors). 

The HAL method has already been used as a tool for a physical analogy
between semantic space and Quantum Theory, where at each word was
associated a given spectrum (in analogy with spectral emission lines
of atoms)  \cite{Wittek}. 

Our method uses the HAL matrices for calculating quantum mechanical
mean values of query observables and combining them in order to calculate
a Bell parameter $S_{query}$.
We carried out our tests in a symmetric matrix obtained by the sum
of the HAL matrix and its transpose (equivalent to run HAL backwards).
This is due to the fact that we did not consider the order in which
words appear in a text.

The tests were carried out on documents in English. The programming
scheme of the algorithm implementation is represented in Figure 3
(Appendix).

\section{Quantum model for Bell Tests using HAL}

In this section we intend to define operators, in analogy with Quantum
Theory, that will give a new possible approach to document queries.
We make the following definitions.

\subsection{Document vector states}

In the $N$ dimensional HAL space each document will have an associated vector.
The vector state of the document is the sum of all the word vectors
$\left|w_{i}\right\rangle$ it contains. Each word vector state is extracted from the lines of the symmetric
HAL matrix. The document vector state is defined as:

\begin{equation}
\left|\Psi\right\rangle =\sum_i^N \left|w_{i}\right\rangle 
\end{equation}

We are now interested in analyzing how two words are connected within
a document, namely word $A$ and word $B$. The two word vectors 
$\left|w_{A}\right\rangle$ and $\left|w_{B}\right\rangle$ define
a plane on the semantic space. We will not consider the part of the
document corresponding to the orthogonal projection with the two chosen words. The resulting normalized state vector
$\left|\psi\right\rangle $ from now on will be the document vector
state. 

To obtain $\left|\psi\right\rangle$ we take the vectors $\left|w_{A}\right\rangle$ and $\left|w_{B}\right\rangle$
and normalize them obtaining two new vectors: $\left|u_{A}\right\rangle$ and $\left|u_{B}\right\rangle$.
Now we apply the Gram-Schmidt orthogonalization process to the non-orthogonal basis
$\{\left|u_{A}\right\rangle,\left|u_{B}\right\rangle\}$,
and doing so, we can obtain two new basis that describe the plane formed by the original vectors
$\left|w_{A}\right\rangle$ and $\left|w_{B}\right\rangle$: the basis
$\left\{ \left|u_{A}\right\rangle ,\left|u_{A}{}_{\bot}\right\rangle \right\} $
and $\left\{ \left|u_{B}\right\rangle ,\left|u_{B}{}_{\bot}\right\rangle \right\} $.
In this way we can parameterize the plane in two ways: in the first we explicit the parallel component of a vector with respect to
$\left|w_{A}\right\rangle$ and in the second we explicit the parallel component of a vector with respect to $\left|w_{B}\right\rangle$
By projecting the vector $\left|\Psi\right\rangle$ on one of these basis, we obtain its projection onto this
plane. Taking this vector and normalizing it gives us the desired vector $\left|\psi\right\rangle$. Explicitly
what we get is:

\begin{equation}
\left|\psi\right\rangle =\alpha\left|u_{A}\right\rangle +\alpha_{\bot}\left|u_{A}{}_{\bot}\right\rangle =\beta\left|u_{B}\right\rangle +\beta_{\bot}\left|u_{B}{}_{\bot}\right\rangle 
\end{equation}

The coefficients $\alpha$, $\alpha_{\perp}$, $\beta$ and $\beta_{\perp}$
are obtained by projecting the state $\left|\Psi\right\rangle $
on both basis vectors and then normalizing to unity. For example for $\alpha$ we have:

\begin{equation}
\alpha=\frac{\left\langle u_{A}\right.\left|\Psi\right\rangle }{\sqrt{\left\langle u_{A}\right.\left|\Psi\right\rangle ^{2}+\left\langle u_{A}{}_{\bot}\right.\left|\Psi\right\rangle ^{2}}} \label{eq:alpha}
\end{equation}

\subsection{Query operators}

We want now to define query operators. The purpose of these operators
is to quantify a query within our formalism. The query operators
$\hat{A}$ and $\hat{B}$ are defined in a way that they attribute
the value $+1$ to the component of the state that corresponds to
the word meaning we are interested in, and $-1$ in
the orthogonal direction. More precisely we will use operators acting
as the spin Pauli matrix $\hat{\sigma}_{z}=\left(\begin{array}{cc}
1 & 0\\
0 & -1
\end{array}\right)$ on their respective decomposition basis. These operators are associated
with observables because they are Hermitian. Explicitly:

\begin{equation}
\begin{array}{ccc}
\hat{A}\left|\psi\right\rangle =\alpha\left|u_{A}\right\rangle -\alpha_{\bot}\left|u_{A}{}_{\bot}\right\rangle  & ,\  & \hat{B}\left|\psi\right\rangle =\beta\left|u_{B}\right\rangle -\beta_{\bot}\left|u_{B}{}_{\bot}\right\rangle \end{array}\label{eq:query-z-operators}
\end{equation}

The expectation values of these operators are calculated in the same way as in
quantum mechanics using the Born rule, for example, the mean value in the context of
document associated to $\left|\psi\right\rangle$ for the query about $A$ is written as usual in quantum mechanics:

\begin{equation}
\left\langle A\right\rangle _{\psi}=\left\langle \psi\right|\hat{A}\left|\psi\right\rangle =\alpha^{2}-\alpha_{\bot}^{2}=2\alpha^{2}-1 
\end{equation}

From this example we see that we can obtain a score for a search related to word $A$. This corresponds to something reasonable for the query score since it increases
with $\alpha$ which is equal to the scalar product between the document vector and
the word vector as shown in Eq. \ref{eq:alpha}. Score values range from $+1$ to $-1$.
 $+1$ is obtained when the document vector is parallel to the query vector,
and $-1$ when it is orthogonal. Following this line
of thought other operators can be defined using, for example, the
Pauli matrix $\hat{\sigma}_{x}=\left(\begin{array}{cc}
0 & 1\\
1 & 0
\end{array}\right)$ . 

For the choice of the query operator $\hat{A}_{x}=\hat{\sigma}_{x}$ in the $\left\{ \left|u_{A}\right\rangle ,\left|u_{A}{}_{\bot}\right\rangle \right\} $ basis we have:
\begin{equation}
\hat{A}_{x}\left(\alpha\left|u_{A}\right\rangle +\alpha_{\bot}\left|u_{A}{}_{\bot}\right\rangle \right)=\alpha_{\bot}\left|u_{A}\right\rangle +\alpha\left|u_{A}{}_{\bot}\right\rangle 
\end{equation}

We see that this operator switches the components of the vector state. This can
be interpreted as a measure of different meaning in the document with
respect to the original direction corresponding to word $A$.

We do not consider the expectation
values for the spin Pauli matrix $\hat{\sigma}_{y}=\left(\begin{array}{cc}
0 & -i\\
i & 0
\end{array}\right)$ due to the fact that here the components of the vector state issued
from the HAL matrix are always real. Possible future generalizations may include
a way of obtaining vector components on the complex plane. 

\subsection{Combining operators and expectation values of two queries}

For technical reasons we choose the basis associated to the word $A$ given above in Eq.
\ref{eq:query-z-operators} and write the operators with respect to
this basis. We can write
the transformation matrix $\hat{M}$ from the $A$ basis to the $B$ basis.
It is easy to see that:
\begin{equation}
\hat{M}=\left(\begin{array}{cc}
\left\langle u_{B}\right|\left.u_{A}\right\rangle  & \left\langle u_{B}\right|\left.u_{A}{}_{\bot}\right\rangle \\
\left\langle u_{B}{}_{\bot}\right|\left.u_{A}\right\rangle  & \left\langle u_{B}{}_{\bot}\right|\left.u_{A}{}_{\bot}\right\rangle 
\end{array}\right)=\left(\begin{array}{cc}
p & \sqrt{1-p^{2}}\\
-\sqrt{1-p^{2}} & p
\end{array}\right)
\end{equation}

By construction, this matrix can be simply expressed by the scalar product
$p=\left\langle u_{B}\right|\left.u_{A}\right\rangle $ which in our case is always
positive and smaller than 1, unless there is a perfect alignment between
the two words, then it is $1$. 

Any operator expressed in its matrix form on the basis associated to the word
$B$ can be written in the basis associated to word $A$ using the transformation
matrix $\hat{M}$. From our previous definition of $\hat{B}$, its matrix
form in the basis associated to word $A$ becomes:

\begin{equation}
\hat{B}=\left(\begin{array}{cc}
2p^{2}-1 & 2p\sqrt{1-p^{2}}\\
2p\sqrt{1-p^{2}} & 1-2p^{2}
\end{array}\right)
\end{equation}
With the two operators expressed in the common basis we can now combine two
query operators and calculate quantum mechanical mean values using the Born rule.
For example for the query of the combination of words $A$ and $B$ in the context of the document represented by
$\Psi$ we will write:
\begin{equation}
\left\langle \hat{A}\hat{B}\right\rangle _{\psi}=\left\langle \psi\right|\hat{A}\hat{B}\left|\psi\right\rangle \label{eq:op-combo}
\end{equation}

\subsection{Bell parameter calculation}

Bell tests are usually a proof of non-separability of the combination of two different
systems. Here we make a connection to this scenario in physics using words and
their meanings within the document.

For purposes of document analysis
we have chosen to take an approach leading to the calculation of a
Bell parameter as defined in Eq. \ref{eq:general-bell}. Concretely
we calculate quantum meanas defined in Eq. \ref{eq:op-combo}, using different query operators which can be
considered as measuring devices, and then define the Bell query parameter:
\begin{equation}
S_{query}=\left|\left\langle \hat{A}\hat{B}_{+}\right\rangle +\left\langle \hat{A}_{x}\hat{B}_{+}\right\rangle \right|+\left|\left\langle \hat{A}\hat{B}_{-}\right\rangle -\left\langle \hat{A}_{x}\hat{B}_{-}\right\rangle \right|\label{eq:bell-query}
\end{equation}
using the following operators
\begin{equation}
\hat{A};\ \hat{A}_{x};\ \hat{B}_{+}=-\frac{\hat{B}+\hat{B}_{x}}{\sqrt{2}};\ \hat{B}_{+}=\frac{\hat{B}-\hat{B}_{x}}{\sqrt{2}}
\end{equation}

Our particular operator choice was inspired from the usual example
that maximizes the violation of the Bell inequalities. All operators have the property of being their
own inverse, that is, their square is the identity (property of the
Pauli matrices) which means that their eigenvalues are $+1$ and $-1$.
With this we can calculate the corresponding parameters considering
different queries among different documents. Two examples are presented
in the next section.

\section{Results and Discussion}

With the formalism presented before we are in a position to apply it to different
documents. We calculated the Bell parameter defined in Eq. \ref{eq:bell-query} using the algorithm
presented in the Appendix.
We will discuss the obtained results in a relevance perspective. In the following
examples all the documents were taken from Wikipedia (see following section).

\subsection{Test on Documents: ``Reagan" and ``Iran"}

As a first application we considered an example originally introduced
by Bruza and Cole  \cite{Bruza-Cole}, which is the query for the word
``Reagan" in the context of ``Iran". If we talk about Reagan alone one
usually associates this with the fact that he was President, but if
we include Iran it will be more likely that we are interested in the
Iran-contra scandal. Four documents were considered which are close
to the query: ``Reagan administration scandals"%
\footnote{http://en.wikipedia.org/wiki/Reagan\_administration\_scandals 
\linebreak (accessed 12/04/2013)%
}, ``Reagan"%
\footnote{http://en.wikipedia.org/wiki/Reagan (accessed 12/04/2013)%
}, ``Iran–Contra affair''%
\footnote{http://en.wikipedia.org/wiki/Iran-Contra\_affair (accessed 12/04/2013)%
} and ``Iran"%
\footnote{http://en.wikipedia.org/wiki/Iran (accessed 12/04/2013)%
}.

\begin{figure}[h]
\begin{centering}

\includegraphics[scale=0.6]{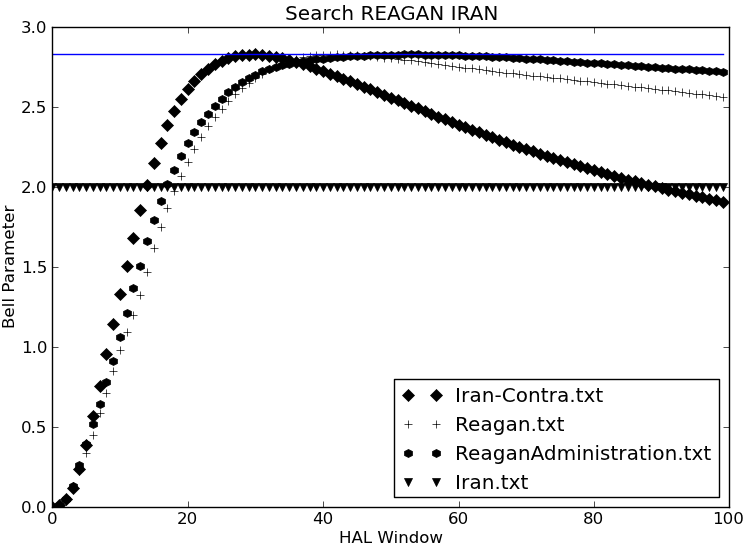}
\par\end{centering}

\caption{\label{fig:Reagan-Iran}Bell parameter for the query of words ``Reagan
- Iran" in four documents: ``Reagan administration scandals", ``Reagan",
``Iran-Contra affair" and ``Iran".}
\end{figure}

We plot the parameters $S_{query}$ defined above in Eq. \ref{eq:bell-query}
as a function of the HAL window length for the query of the words ``Reagan"
and ``Iran". The results are shown in Figure \ref{fig:Reagan-Iran}. 

The considered HAL window starts at the beginning of the document, in the first word, 
and will run through all the words until the end of the text. When finished we start the same process again
 with a new window length (increasing the length by one unit). This is done for all window lengths we desire to analyze.

There is clearly a common behavior for the three queries in the documents
(with just one exception): the parameter starts from zero and increases
until it reaches a maximum, never crossing the Tsirelson's bound $2\sqrt{2}$
, but getting very close to it, and then drops again. This suggests
that each document, given a two word query, has an optimal HAL window
size that maximizes the parameter $S_{query}$.

 For the query of ``Iran
- Reagan", among the four documents, it is predictable that the document
that is more closely related to this query is the ``Iran-Contra affair",
followed by the documents: ``Reagan administration scandals" and ``Reagan",
with an expected greater relevance for the first. The least related
document should be ``Iran". 

At first sight it may appear that since
we are looking for ``Reagan - Iran", the documents ``Reagan" and ``Iran"
should appear on the same level in the search. However in general,
the meaning ``Reagan" has less importance in the context ``Iran" (because
the common concept ``Iran" includes its history, culture, geographical
situation, etc.) than ``Iran" in the context of ``Reagan". In Figure
\ref{fig:Reagan-Iran} we also observe that this is basically the
order in which the peaks appear.

The document regarding ``Iran" always gives a constant value of $S_{query}$ equal to $2$.
This fact is easily explained in the framework of our model. In fact
it is not hard to see that when we do a two word query in which one
of these words is not present in the document, the result for $S_{query}$
is always $2$. Besides if neither of the two words is present the
result is always zero. Let us now consider the other three documents.

The first peak appears using a window length around $l=30$. The $S_{query}$
curve peaks before this value in the document of ``Iran-Contra affair".
The other two documents cannot be clearly distinguished. This corresponds
to our previous prediction. In fact, it makes sense that the ``sooner"
a peak appears the less interaction, in the sense of window length,
we have to consider to get higher correlation between the two words.
Bearing this in mind, the document ``Iran-Contra affair" is clearly
the one selected by the model. The other two documents ("Reagan" and
``Reagan administration scandals") are not clearly distinguishable.
On one side the peak of the ``Reagan" document appears first, but the
curve for ``Reagan administration scandals" has a bigger extension
close to the Tsirelson's bound $2\sqrt{2}$, meaning high correlation
for several window sizes, which can also be a clue for some strong
correlation between words.

\subsection{Test on Documents about ``Orange"}

\begin{figure}
\begin{centering}
\includegraphics[scale=0.4]{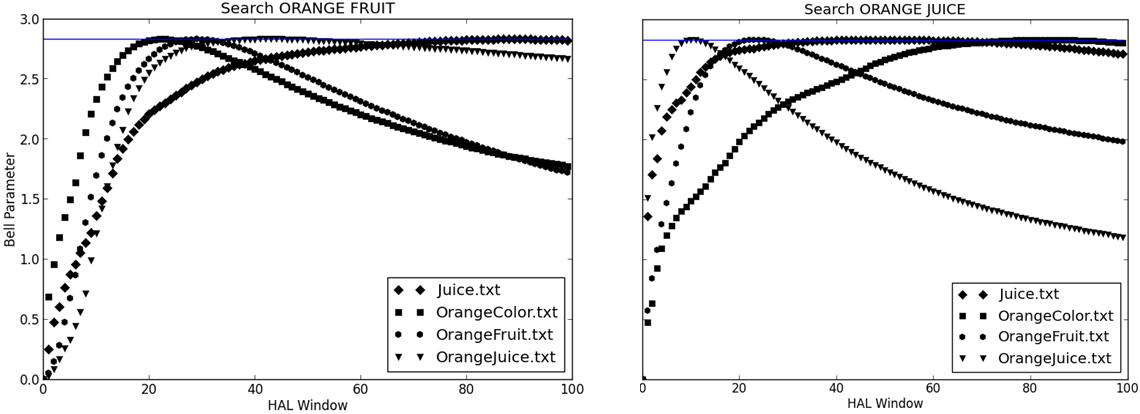}
\par\end{centering}

\caption{\label{fig:Orange}Bell parameter for the query on the words ``Orange
- Fruit" and ``Orange - Juice" in four different documents: ``Juice",
``Orange (Color)", ``Orange (Fruit)" and ``Orange Juice".}
\end{figure}

The second case considered concerns the polysemy of the word ``orange"
and associated concepts. In this example we are interested in the
ambiguity between the meanings color and fruit. We also associate
the concept of juice. The documents considered were: Orange (Colour)%
\footnote{http://en.wikipedia.org/wiki/Orange\_(colour) (accessed 12/04/2013)%
}, Orange (Fruit)%
\footnote{http://en.wikipedia.org/wiki/Orange\_(fruit) (accessed 12/04/2013)%
}, Orange Juice%
\footnote{http://en.wikipedia.org/wiki/Orange\_juice (accessed 12/04/2013)%
} and Juice%
\footnote{http://en.wikipedia.org/wiki/Juice (accessed 12/04/2013)%
}. Two queries are considered: ``Orange Fruit" and ``Orange Juice". The
results are presented in Figure \ref{fig:Orange}.

The query ``Orange - Fruit" presents the first peak around $l=22$
for the document ``Orange color", the second in $l=29$ for the document
``Orange fruit", then for $l=40$ the document ``Orange juice" and very
far away the document ``Juice". It is interesting to note that, even
though the peaks are close, the peak of the curve ``Orange - Color"
appears before the one for ``Orange - Fruit". This may be the suggestion
of a strong correlation between the origin of the name of the color
and the name of the fruit. The poor correlation of the general term
``Juice" with the specific query ``Orange - Fruit" is very clear on
the graph according to this criterion. 

Finally, the last query was ``Orange - Juice". Again, here, we recover
precisely the order that we would expect for the documents: ``Orange
juice", ``Orange Fruit", ``Juice" and ``Orange Color". It is worth noticing
that in the latter case the peak corresponds to a window length range
considered to be optimal for implementation of HAL ($l=10$) which
may indicate an even a stronger correlation between the words.

\section{Conclusion and Perspectives}

In this work, we presented a novel search experiment based on the
Bell parameter extraction in semantic space using the HAL method. 

The semantic vectors in HAL are representations that are essentially
measures of context. The HAL method has already been used for the
analogies with Quantum Theory by Bruza and Woods  \cite{Bruza-Woods} for activating
associations of concepts and by Wittek and Dar\'{a}ny  \cite{Wittek} for
extracting spectral content from the semantic space. HAL shows high
potentiality because it is a simple way to build a semantic space
with a measure that is independent of user judgment. 

The main feature of Quantum Theory explored in this work is the violation
of the Bell inequalities which can be related to entanglement and
non-locality, impossible at a classical level. The results show Bell
inequality violation up to the maximal value of $S_{Bell}=2\sqrt{2}$,
(the Tsirelson's bound). 

In our model each document is associated
to a two dimensional Hilbert space (dependent on the search we are
interested in), and queries are observables acting on it. A Bell parameter
is then defined. 

We found that the Bell parameter is strongly dependent
on the HAL window size. From our results it is suggested that for
this kind of model there is an optimal window size that maximizes
the Bell parameter. This is reminiscent of what was also noticed by
Bruza and Woods  \cite{Bruza-Woods}: if the window size is set too
large spurious co-occurrence associations are represented in the matrix
and, if the window size is too small, relevant associations may be
missed. In this model we see that too large windows may also dilute
connections between associated words. 
Only one document, the one that did not present one of the words of
the query, did not violate the Bell inequality. In general, a pattern
of ``early" appearance of the peak (smallest window sizes) seems to
be related to the relevance of the document for the search. 

In a near
future other measures of quantum properties (as proper measures of
entanglement) will also serve to make a better comparison between
the results issued from the standard information retrieval methods.

It is not clear how to interpret the Bell inequality violation here
and what is the meaning of the optimal length that maximizes the Bell
parameter. 
Can correlation and entanglement give a measure of query relevance?
Experiments and systematic comparisons with other methods used in
IR, such as Latent Semantic Analysis (LSA) and the ranking method
Okapi BM25, could give further indications. 

An important technical
point is that we introduced a new tool which has connections with
the Quantum Theory: query observables. Here we made a practical choice
similar to the spin Pauli matrices, but we think that it should be
possible, after much experimentation on different documents, to introduce
new families of query observables adapted to different purposes and
contexts. In the domain of IR many concepts are introduced to define,
for example, opinion-like queries in social networks  \cite{Youssef}.
Efforts are also being made in order to diversify query results of ambiguous queries,
for example concepts such as sentiment diversification to identify positive, negative
and neutral sentiments about the search topic can be used  \cite{Kacimi}.

\section*{Acknowledgements}
We would like to thank the students of SUPELEC Fabien d’ANGELO and
Sixte BOISS\'{E} for helping on the implementation and test the HAL algorithm.

\appendix
\section*{Appendix}\FloatBarrier

\begin{figure}[h]
\begin{centering}
\includegraphics[scale=0.8]{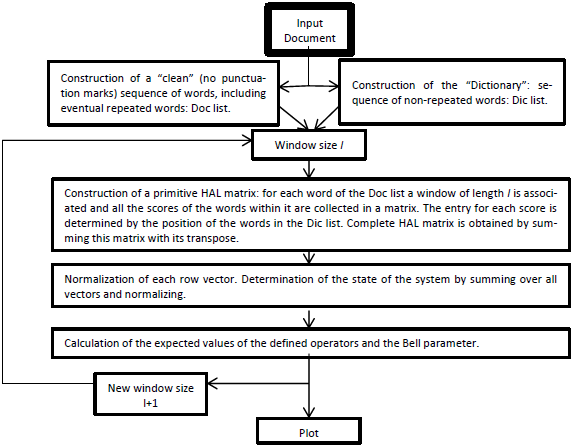}
\par\end{centering}
\caption{Flow diagram of the Quantum HAL algorithm described.}
\end{figure}

The algorithm was implemented using Python programming language along
with the string module and pylab. All words are considered and simple
plurals (constructed by adding an ``s") are treated as if singular
words. Lower and upper case letters are not distinguished, which means
that if two words differ from each other by changes on the case, they
are considered equivalent. Even if words have the same origin, they
are treated differently (for example ``battle" and ``battling" are distinct).

\end{document}